# Plasmon-Exciton Coupling and Dephasing in Hybrid Au Nanostructure/J-Aggregate Systems

Janak Bhandari, Robert Catuto, Zhumin Zhang, Bradley D. Smith, Hsing-Ta Chen and Gregory V. Hartland[1*]

Department of Chemistry and Biochemistry
University of Notre Dame
Notre Dame, IN 46556
United States

**Abstract:** The coupling between propagating surface plasmon polaritons (SPPs) in Au nanostructures and the exciton transitions of cyanine dye J-aggregates has been examined using leakage radiation microscopy. Real space images of the nanostructures give the propagation lengths of the leaky SPP modes, and Fourier space images yield their dispersion curves. The dispersion curves show an avoided crossing when the structures are coated with J-aggregates, with a Rabi splitting of approximately 30 meV. The lifetimes of the coupled states were calculated by combining the measured propagation lengths with the group velocities obtained from the dispersion curves. The lifetimes decrease from ~50 fs for the bare Au nanostructures, to ~10 fs in the avoided crossing region for the coupled J-aggregate/Au nanostructure system. Analytical Holstein-Tavis-Cummings model calculations and finite element simulations of the coupled system show that the decrease in lifetime is primarily due to energy dissipation into dark states associated with the J-aggregates.

---

[1*] Corresponding Author; e-mail: ghartlan@nd.edu

1. **Introduction**

The formation of hybrid light–matter states, known as polaritons, is of significant recent interest in the nano-photonics and quantum-materials communities.[1-6] Polaritons arise when the rate of energy exchange between a photonic mode and an optical transition of a semiconductor or molecular system exceeds their respective decay rates.[7-8] This creates hybrid states that share both photonic and material excitation properties. Polaritons are typically characterized by their Rabi splitting, which can be measured from dispersion curves for the coupled light–matter systems.[1, 7-11] Polaritons enable long range energy transfer[12] and nonlinear optical responses within subwavelength volumes,[13] and have been used to study of phenomena such as Bose–Einstein condensation.[2, 14]

In the past decade polaritons formed through the coupling of surface plasmons with excitons have been extensively investigated.[1, 9-11] An important attribute of surface plasmons is that they confine electromagnetic fields to dimensions far below the diffraction limit, enhancing the interaction with excitons and enabling strong coupling without the need for an external cavity.[1, 9-11] Experiments have been performed on localized surface plasmon resonances,[9, 15] surface lattice resonances,[16-17] and propagating surface plasmon polaritons (SPPs)[18-20] which are the focus of this work. SPPs are guided waves that move along the interface between an extended metal structure (such as a surface or nanowire) and a dielectric material.[21-22] Of the different excitonic systems that have been examined, J-aggregates of organic dyes are probably the most effective for forming polaritons, as they possess narrow absorption bands and large oscillator strengths.[9, 15, 18-20] Rabi splittings of hundreds of meV have been observed for plasmons coupled to J-aggregates,[5, 10, 15, 18-20, 23] placing these systems within the strong coupling regime.[7-8]

There has been extensive research into the spectral and dispersion properties of plasmon-exciton polaritons, however, less information is available about their dynamics.[8, 24] One reason for this is that plasmons have very short lifetimes,[25] so direct time resolved measurements are difficult.[24, 26] Light scattering spectra from single plasmonic particles can yield lifetime information,[25, 27] however, these spectra are challenging to interpret for coupled plasmon-exciton systems.[15, 23] To the best of our knowledge there are only a few quantitative measurements of polariton lifetimes for plasmonic systems, all by two-dimensional electronic spectroscopy.[28-30] The polariton lifetime dictates how long the hybrid states preserve their coherence.[8]

Understanding how the coherence time varies with the identity of the excitonic species and coupling strength is important for designing polaritonic devices, where performance hinges on balancing field confinement and dissipation.[8, 13]

Here, we report leakage radiation microscopy experiments for J-aggregates coupled to SPPs of gold nanostructures.[25, 31] Leakage radiation microscopy was used for these experiments because it can provide information about the dephasing time of the SPP states. The measured $T_1$ times for the coupled system are approximately 10 fs, much shorter than the lifetime of the SPPs in the bare Au nanostructures (ca. 50 fs).[25, 31-32] This is somewhat counter-intuitive. The simple coupled oscillator model for mixed states predicts that the dephasing should be an average of the damping rates of the two oscillators.[7-8] The width of the J-aggregate absorption band is on-the-order-of 30 meV. This linewidth has contributions from a variety of sources, including electron-phonon coupling, dephasing and inhomogeneous broadening,[33-34] which means that the $T_1$ lifetime must be longer than 25 fs.[25] This implies that the coupled states should have lifetimes > 33 fs, significantly longer than the measured lifetimes. An analytical model for the SPP-exciton system based on the Holstein-Tavis-Cummings Hamiltonian reveals that the decrease in lifetime arises from coupling between the SPP mode and dark states associated with the exciton. This is corroborated by finite element simulations, which show that the lifetimes of the polariton states created in this system suffer additional energy losses by resistive heating in the dye layer as well as radiation damping,[25, 31-32, 35-36] with resistive heating being the main effect in the avoided crossing region.

2. Methods

The gold nanostripes were fabricated on #1.5 borosilicate glass substrates using photolithography, followed by the physical vapor deposition and liftoff. The structures for the present study were 2.6 μm wide, 100 μm long, with a thickness of 50 nm. J-aggregates of the pentamethine cyanine dye, ZZ683, in a poly (vinyl alcohol) (PVA) solution were spin coated onto the substrate to create hybrid dye-Au nanostripe systems. J-aggregates were selected for this study due to their strong excitonic absorption and narrow linewidths, ideal for achieving strong coupling with plasmonic modes. The specific cyanine dye for this study was chosen because it forms J-aggregates in the near-IR region,[37] which is important for leakage radiation studies of the Au nanostripes. The thickness of the dye-PVA layer was estimated to be approximately 20 nm with a refractive index of 1.5 by ellipsometry measurements (Gaertner L117 Ellipsometer).

A schematic of the experimental setup is presented in the Figure S1 of the Supporting Information. Briefly, the samples were placed on an inverted optical microscope (Olympus IX-71) and optically excited across a range of wavelengths from 730 nm to 820 nm with the output of a white light supercontinuum laser (NKT Photonics SuperK COMPACT).[35-36] Specific wavelengths were selected using a series of 10 nm bandpass interference filters (Thorlabs FBHXXX-10). The laser beam was tightly focused onto the end of the nanostripes using a high numerical aperture (NA) oil immersion objective lens (Olympus UPlanFL N, 100x, NA=1.3) to launch the SPPs.[31] The overlap between the laser focus and the nanostripes was achieved with a Mad City Laboratories MCL-uS motorized stage. The scattered light from the sample was collected using the same objective and was directed to a CMOS camera (Thorlabs, Kiralux CC895MU). Real space and Fourier space images of the nanostripe SPP modes were collected for all wavelengths.[38-43] The SPP propagation lengths $L_{SPP}$ were obtained from the real space images, and the SPP wavevectors $k_{SPP}$ were obtained from the Fourier space images. The absorption spectrum of the near-IR cyanine dye-PVA film was recorded with a Jasco UV-670 UV-vis spectrophotometer.

Finite element simulations of the nanostripes SPP modes were carried out in COMSOL Multphysics (ver 5.3a). A two-dimensional mode analysis calculation was performed in an "Electromagnetic Waves, Frequency Domain" simulation.[43-44] In our model a 3 nm Ti adhesion layer was included between the nanostripe and the glass substrate to match the experimental system. Values for the refractive index for Ti were taken from COMSOL's library of materials, and the refractive index data for Au was taken from Ref. [45]. The refractive index of glass was set to 1.45 (the value from COMSOL for silica glass). The mode analysis calculation yields the complex effective mode index for the system $n_{eff} - i\frac{\alpha}{k_0}$, where the real part gives the SPP wavevector $k_{SPP} = n_{eff} k_0$, and the imaginary part gives the attenuation constant α, which is related to the propagation length by $L_{SPP} = \frac{1}{2\alpha}$.[38, 40-41, 44] The values of α obtained from the simulations can be separated into contributions from radiation damping, energy transfer to the dye layer, and resistive heat in the Au (Landau damping) by calculating the power losses for these different processes.[25, 35-36, 43] Specifically, in our two-dimensional simulations the relative power dissipated by radiation was calculated from the line integral of the time-averaged Poynting vector $\langle\vec{S}\rangle$ over a circle that encloses the nanostripe: $\oint \vec{n}.\langle\vec{S}\rangle dl$ where $\vec{n}$ is the outward normal unit vector.[43] The powers dissipated by resistive heating in the metal or dye/polymer layer were calculated by $\iint Q_{rh} dA$ where $Q_{rh}$ is the resistive losses in the material and the integral is over the metal or dye/polymer domains.[35-36] The different contributions to the SPP attenuation were then calculated by $\alpha_i = \eta_i \alpha_{tot}$ where $\alpha_{tot}$ is the total attenuation from the mode analysis calculation, and $\eta_i$ are the relative power losses for radiation damping ($\eta_{rad}$) and resistive heating ($\eta_{Au}$ and $\eta_{dye}$).[35-36, 43]

### 3. Results and Discussion

Metal nanostripes on a glass surface have two SPP modes: a bound mode that propagates at the metal-glass interface, and a leaky mode at the metal-air interface. The wavevector of the bound mode is too large to couple to light, but the leaky mode can couple to photons in the glass.[43] Figure 1(a) shows a momentum matching diagram for coupling between the leaky SPP mode and photons in the glass substrate. Momentum matching occurs when the SPP wavevector $k_{SPP} = nk_0 \sin\sin\theta \cos\cos\varphi$, where $k_0$ is the momentum of free space photons and $n$ is the refractive index of the substrate. Using the coordinate system shown in Figure 1(a), this gives a line in the Fourier space image at $\left(\frac{k_x}{k_0}, \frac{k_y}{k_0}\right) = \left(\frac{k_{SPP}}{k_0}, \frac{k_{SPP}\tan\tan\varphi}{k_0}\right)$.[43] Note that the maximum wavevector that can be collected by the optical system is determined by the numerical aperture of the objective. Figure 1(b) shows an example Fourier space images from a Au nanostripe used in these experiments. The inner circle in the Fourier space image corresponds to the condition for total internal reflection $\frac{k}{k_0} = 1$, and the outer circle is $\frac{k}{k_0} = NA$. These two features calibrate the wavevector scale in the Fourier space images.[43] Figure 1(c) shows a real space image of the leaky SPP mode. The SPP propagation lengths were measured by integrating the scattered light over the width of the nanostripe, and fitting the resulting intensity profile to an exponential decay $I(x) = e^{-\frac{x}{L_{SPP}}}$, see Figure 1(d). SPP dephasing can be characterized by either attenuation constants/propagation lengths or lifetimes. The lifetime is more intuitive, and allows comparison to other systems. The SPP lifetimes are given by $T_1 = \frac{L_{SPP}}{v_g}$,[25, 31, 35-36, 43] where $v_g$ is the group velocity. The group velocities were measured by recording Fourier space images over a range of frequencies to construct a frequency versus wavevector dispersion curve, and using $v_g = \frac{\partial \omega}{\partial k}$.[46]

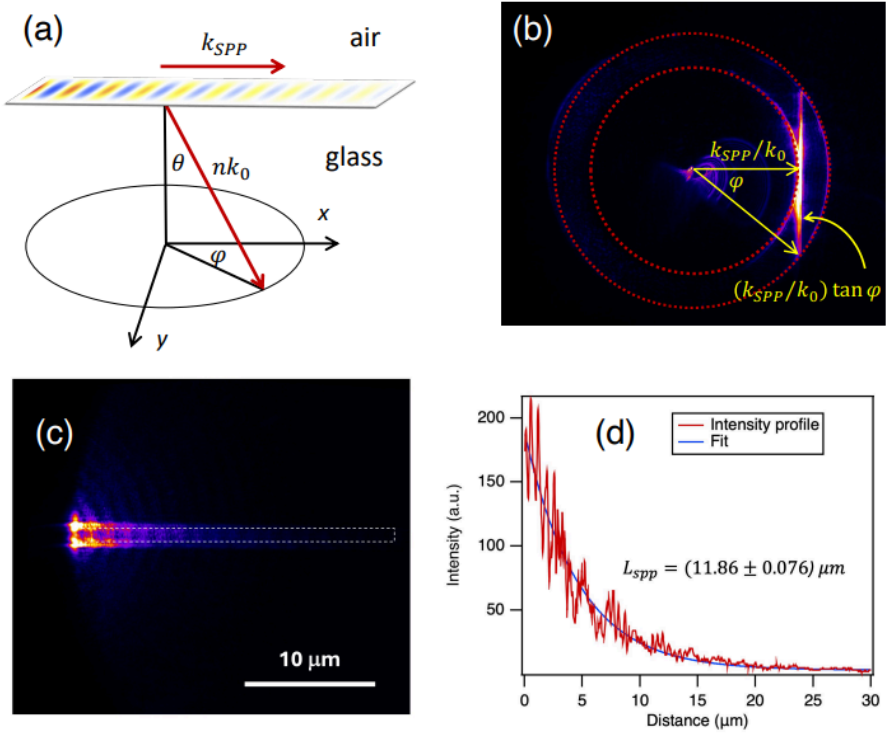

**Figure 1:** (a) Momentum matching diagram for the leaky SPP mode (wavevector $k_{SPP}$) coupling to photons in the glass substrate (wavevector $nk_0$ where $n$ is the refractive index of the glass). (b) Example Fourier space image of the leaky SPP mode of a typical Au nanostripe used in our experiments. The inner circle corresponds to $\frac{k}{k_0} = 1$, and the outer circle to $\frac{k}{k_0} = NA$, where $NA$ is the numerical aperture of the objective. (c) Example real space image. (d) Line profile obtained by integrating the scattered light over the width of the nanostripe. The black line shows an exponential fit to the data.

The excitonic system used for these experiments was a pentamethine cyanine dye, denoted as ZZ683, which can be induced to form J-aggregates by controlling solution conditions.[47-50] Figure 2(a) shows absorption spectra of the dye as monomer in methanol, J-aggregate form in phosphate buffered saline solution, and spin coated J-aggregate on a glass substrate. Spin coating broadens the J-aggregate band, indicating a reduction in the exciton delocalization length.[51] Figure 2(b) shows the frequency versus wavevector dispersion curves for the leaky SPP modes of bare Au nanostripes (green lines) and J-aggregate coated Au nanostripes (blue lines). The red line is the dispersion curve for free space photons ($\omega = c_0 k_0$), and the dotted black line shows the absorption spectrum of the J-aggregates on glass. The results from individual nanostripes are shown as light green/blue lines, and the averaged data as dark green/blue lines. Eight different nanostripes were examined for the coated system, and three different nanostripes for the bare system (the wavevectors and propagation lengths are typically very consistent for bare Au nanostripes, so a large number of measurements are not needed). The dispersion curves show a discontinuity at the maximum of the J-aggregate band. The Rabi splitting between the upper and the lower polariton branches was estimated to be approximately 30 meV (see the Supporting Information). This is similar to the linewidth of the J-aggregates spectrum, placing the system in an intermediate coupling regime: strong enough to produce an avoided crossing in the dispersion curve, but not in the strong coupling limit where the Rabi frequency greatly exceeds the SPP and exciton dephasing rates.[7-8] The propagation lengths versus wavelength for bare and J-aggregate coated Au nanostripes are presented in Figure 2(c) (same color scheme as Fig. 2(b)). It is important to note that the SPP propagation length decreases near the J-aggregate excitonic frequency in the coupled system, which implies that the interaction with the exciton transition causes increased SPP attenuation and, thus, decreased lifetimes.

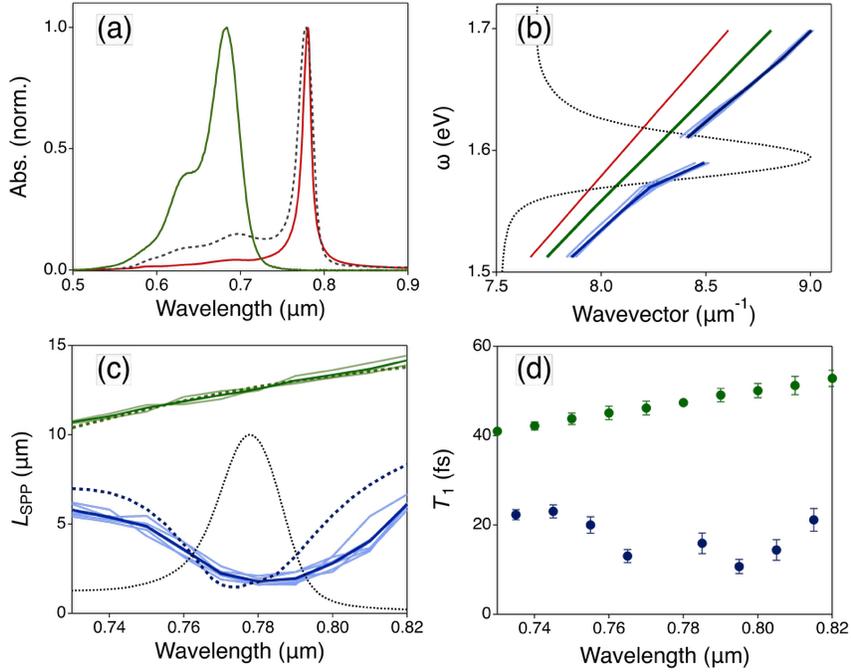

**Figure 2:** (a) Absorption spectra of the pentamethine cyanine dye, ZZ683, in monomer form in methanol solution (green line), and as J-aggregates in phosphate buffered saline solution (red line). The dashed black line corresponds to the J-aggregates spin coated onto a glass surface. (b) Dispersion curve for bare Au nanostripes (green lines) and J-aggregate coated nanostripes (blue lines). Results from individual nanostripes are shown as light green/blue lines, and the averaged data as dark green/blue lines. The red line is the light line ($\omega = c_0 k_0$), and the dotted black line shows the absorption spectrum of the J-aggregates on glass. (c) Propagation lengths for bare and J-aggregate coated nanostripes, and the absorption spectrum of the J-aggregates on glass, same color scheme as panel (b). The dashed lines are the propagation lengths calculated from the finite element simulations. Note that the cyanine dye absorption spectra in panels (b) and (c) are simply scaled to fit the figure. (d) Experimental lifetime versus wavelength for bare nanostripes (green symbols) and J-aggregate coated Au nanostripes (blue symbols). Error bars are 95% confidence limits that were determined from errors in both $L_{SPP}$ and $v_g$.

Figure 2(d) shows the $T_1$ lifetimes of the bare and dye coated nanostripes obtained by dividing the propagation lengths (Fig. 2(c)) by the group velocity (determined from Fig. 2(b)). In this analysis the group velocities for the bare nanostripes were obtained by fitting the dispersion curve data to a polynomial function and taking the derivative, allowing $v_g$ and thus $T_1$ to be calculated at each wavelength where $L_{SPP}$ was measured. For the coated nanostripes, the group velocity was determined point-by-point from the averaged wavevector data by $v_g = \frac{\Delta \omega}{\Delta k}$. This yields $v_g$ values in between the experimental data points, which requires interpolated values of $L_{SPP}$ to be used for the $T_1 = \frac{L_{SPP}}{v_g}$ calculations. The SPP lifetime is significantly shorter for the J-aggregate coated nanostripes for all wavelengths, as expected from the propagation length data in Fig. 1(c). Importantly, there is a significant decrease in the lifetime near the J-aggregate exciton resonance. To understand the origins of these lifetime changes, analytic Holstein-Tavis-Cummings (HTC) model calculations were performed, as well as finite element simulations for the bare and J-aggregate coated gold nanostripes.

In the analytic calculations a minimal HTC model Hamiltonian was used composed of $N$ molecular excitons coupled to their individual phonon bath, interacting with a single SPP mode.[52] The SPP mode is modeled as a cavity photon of frequency $\omega_c$ and loss rate $\Gamma_c$, and we focus on the single photon excitation manifold ($|0\rangle$ and $|1\rangle$ as photon states). The electronic excitation of the dye molecule is modeled as an effective two-level system ($|g_n\rangle$ and $|x_n\rangle$ for the ground and excited states of molecule $n$) with the transition frequency $\omega_0$ and exciton relaxation rate $\Gamma_x$. Each dye molecule in the J-aggregate is assumed to have an identical light-matter coupling strength $g_c$, so that the effective coupling to the bright excitonic state ($|B\rangle = \sum_n \frac{1}{\sqrt{N}} |x_n\rangle$) is $\sqrt{N} g_c$. The polaritonic states can be expressed as a linear superposition of the bright excitonic state and cavity photon excitation: $|P_+\rangle = \cos \Theta |B\rangle|0\rangle + \sin \Theta |G\rangle|1\rangle$ and $|P_-\rangle = -\sin \Theta |B\rangle|0\rangle + \cos \Theta |G\rangle|1\rangle$ for the upper/lower polaritons (UP/LP), respectively. Here $|G\rangle$ is the ground state, and the mixing angle is given by $\Theta = \frac{1}{2}\arctan[\frac{2\sqrt{N}g_c}{\omega_c - \omega_0}]$.

The nuclear degrees of freedom of the dye molecules are modeled as an independent, non-interacting harmonic bath with a co-linear coupling to the electronic exciton:

$V_{sb} = \sum_n |x_n\rangle\langle x_n| \times \sum_\alpha c_\alpha R_{n\alpha}$, where $R_{n\alpha}$ is the α-th bath mode coordinate of molecule $n$.[9-10] The system-bath coupling strength $c_\alpha$ is characterized by the spectral density $J(\omega)$, taking the Drude-Lorentz form centered at $\omega_0$

$$J(\omega) = \frac{\pi}{2}\sum_\alpha \frac{c_\alpha^2}{\omega_\alpha}\delta(\omega - \omega_\alpha) = \frac{2\lambda\omega_b\omega}{(\omega-\omega_0)^2+\omega_b^2} \tag{1}$$

where $\omega_b$ is the characteristic bath frequency and $\lambda$ is the reorganization energy, which represents the system-bath coupling strength. The system-bath coupling induces inter-branch scattering to the dark states, which are purely excitonic states that do not involve the polaritonic states, leading to additional decay of the polaritonic states.[53-54] The total decay rate can be expressed as a weighted sum of three decay channels:

$$\Gamma_{tot} = |C|^2\Gamma_c + |X|^2\Gamma_x + \Gamma_d \tag{2}$$

where the Hopfield coefficients are $|C|^2 = \sin^2\Theta$ and $|X|^2 = \cos^2\Theta$ for the UP branch, $|C|^2\cos^2$ and $|X|^2\sin^2$ for the LP branch, and $\Gamma_d$ is the decay to the dark states. In the presence of the electron-nuclear coupling, the scattering to the dark states can be estimated by Fermi's golden rule: $\Gamma_d = |X|^2 J(\omega_c)$.[53-54] For the exciton decay, we estimate $\Gamma_x^{-1}$ from the absorption spectrum of the J-aggregates (Figure 1(a)). This value of $\Gamma_x$ is less than the experimental linewidth, with the additional broadening coming from vibrational relaxation.[34] The reorganization energy and the bath characteristic frequency for the spectral density were chosen to be $\lambda = 0.001\, fs^{-1}$ and $\omega_b = 0.026\, \mu m^{-1}$ from fitting the upper/lower polariton lifetimes for $\omega_c > \omega_0$ and $\omega_c < \omega_0$, respectively (see Supporting Information). Note that we do not fit the spectra for the uncoupled J-aggregates, as the presence of the metal surface could change the dye relaxation.

The dephasing rates from Equation (2) are compared to the experimental measurements in Fig. 3(a). The analytic result qualitatively agrees with the experimental data. Note that without the dark-state decay, the polariton decay rate is a linear combination of $\Gamma_c$ and $\Gamma_x$, and cannot capture the enhanced relaxation of the coupled system at the resonance $\omega_c \approx \omega_0$. This implies that the dark-state decay channel dominates the overall relaxation of the coupled system. Also, using different values for the exciton decay $\Gamma_x$ does not change the form of the $\Gamma_{tot}$ versus wavelength curve, see the Supporting Information. Thus, the observation of increased relaxation due to dark-states is not sensitive to the exact values of $\Gamma_x$, $\lambda$ and $\omega_b$ used in the calculations.

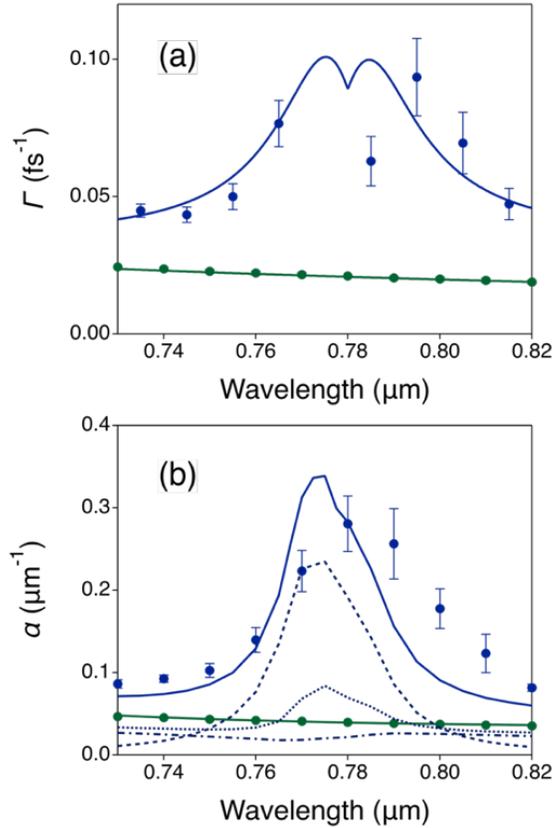

**Figure 3:** (a) Dephasing rates $\Gamma$ estimated by the analytical Holstein-Tavis-Cummings model for bare Au (green lines) and the coupled J-aggregate Au nanostripe system (blue lines). The symbols are the experimental data (Error bars = 95% confidence limits). (b) Attenuation constant $\alpha$ versus wavelength for J-aggregate coated Au nanostripes from the finite element simulations. Same color scheme as panel (a). The dashed and dashed-dotted lines are the calculated attenuation from resistive heating in the dye layer and the Au, respectively, and the dotted line shows the radiation damping contribution. For the bare Au only the total calculated attenuation is presented.

The effect of J-aggregates on plasmon dephasing was also be analyzed by finite element simulations. The dye layer is modelled as a uniform 20 nm film that coats the nanostripes, with a complex dielectric constant of $\epsilon_{film}(\omega) = x\epsilon_{dye}(\omega) + (1-x)\epsilon_m$. The frequency and width of the Lorentz oscillator $\epsilon_{dye}(\omega)$ were adjusted to match the spectrum of the J-aggregate film,[55] and the amplitude of $\epsilon_{dye}(\omega)$, mixing fraction $x$ and dielectric constant of the medium $\epsilon_m$ were adjusted to match the dispersion curve and the attenuation constant at a single wavelength (taken to be 0.78 $\mu m$) for the Au nanostripe/J-aggregate system. More details about the simulations and Lorentz oscillator model are given in the Supporting Information. The attenuation constants α determined from the simulations are compared to the experimental results in Fig. 3(b). Note that we have chosen to present attenuation here, rather than propagation length or lifetime, as α can be directly compared to Γ in Fig. 3(a), and can be also be separated into the different damping contributions (radiation damping, and resistive heating in the J-aggregate and gold domains) in the finite element simulations.[25, 46] The solid line in Fig. 3(b) shows the total calculated attenuation, which is in good agreement with the experimental data. The dashed line is the attenuation from resistive heating in the dye layer, the dotted line is the contribution from radiation damping, and the dashed-dotted line is resistive heating in Au nanostripe.[25, 35-36, 43] These results show that the increased attenuation for the J-aggregate coated nanostripes at the exciton resonance primarily arises from energy dissipation in the dye. There is also a slight increase in radiation damping, that comes from the change in the refractive index of the dye film at the resonance. This is consistent with our previous studies of dye and semiconductor coated Au nanostripes.[35-36]

Both the analytic model and the finite element simulations show that the polariton lifetimes are limited by energy dissipation into non-radiative states of the dye, and that the simple coupled oscillator model does not properly capture the behavior of this system.[7-8] This is a consequence of the strong optical absorption of the J-aggregates. It is important to note that the decrease in lifetime/increased attenuation observed in this work is likely to be a general feature of coupled plasmon-exciton states. Creating these states requires strong exciton transitions, and these transitions are always associated with strong absorption.[47-50] This effect is important for applications: decay into the dark states associated with the exciton limits the propagation lengths that can be achieved for the coupled SPP-exciton system.

4. **Summary and Conclusions**

Hybrid plasmon-exciton states have been created by coating Au nanostripes with J-aggregates of a pentamethine cyanine dye (ZZ683).[37, 47-50] The states were interrogated by single nanostructure light scattering measurements, which probe the leaky SPP modes of the nanostripes.[25, 31-32, 35-36, 43] The dispersion curve for the coupled Au nanostripe/J-aggregate system show an avoided crossing, with a corresponding Rabi frequency of approximately 30 meV. The lifetimes of the plasmon-exciton states were found to be on the order of 10 fs, significantly shorter than the lifetime of the leaky SPP mode of the bare nanostripes. Analytic HTC model calculations and finite element simulations show that the increased attenuation arises from energy dissipation in dark states of the J-aggregates. Similar effects have been observed for Au nanostripes coupled to non-J-aggregate forming dyes and semiconductor systems.[35-36]

The coupled states in our experiments have propagation lengths of several microns. This is longer than the typical propagation lengths of excitons in room temperature J-aggregates,[56-59] but smaller than what has been observed for excitons of organic materials coupled to surface plasmons in nanoparticle arrays.[4, 16-17, 59-60] However, it is also important to note that the leakage radiation measurements used here are only sensitive to coherent energy transport (propagation before dephasing of the polariton states). Any subsequent diffusional motion of the excitons is not detected in our experiments. Finally, the Au nanostructure/J-aggregate system studied here has a relatively modest Rabi frequency. Work is currently underway to find strongly coupled plasmon-exciton systems by investigating different dyes and metals (i.e., Ag rather than Au).

**Acknowledgments:** This work was supported by a grant from the National Science Foundation (CHE-2304905), and the nanostructures used in the study were fabricated in the Notre Dame Nanofabrication Facility. ZZ and BDS acknowledge support by NIH grant R35GM136212.

**Supporting Information:** Description of the optical system used for leakage radiation microscopy measurements; synthesis and characterization of the cyanine dye; details of the finite element simulations; determination of parameters for the analytic HTC model calculations.

**ORCID ID Numbers:**

Janak Bhandari: 0000-0001-8842-3259

Robert Catuto: 0009-0004-4252-6489

Zhumin Zhang: 0000-0002-9780-5571

Bradley D. Smith: 0000-0003-4120-3210

Hsing-Ta Chen: 0000-0002-6619-1861

Gregory V. Hartland: 0000-0002-8650-6891